\documentclass{mn2e}
\usepackage{color}
\usepackage{graphicx}
\usepackage{dcolumn}
\usepackage{bm}

\begin{document}

\title[The SMBH mass density]
{The local supermassive black hole mass density: 
corrections for dependencies on the Hubble constant}

\author[Graham \& Driver]
{Alister W.\ Graham$^{1}$\thanks{AGraham@astro.swin.edu.au}
 and Simon P.\ Driver$^{2}$\\
$^1$Centre for Astrophysics and Supercomputing, Swinburne University
of Technology, Hawthorn, Victoria 3122, Australia\\
$^2$SUPA\thanks{Scottish Universities Physics Alliance (SUPA)}, 
School of Physics \& Astronomy, University of St Andrews, 
North Haugh, St Andrews, Fife, KY16 9SS, UK
}

\date{Received 2006 Jan 01; Accepted 2006 December 31}
\pubyear{2006} \volume{000}
\pagerange{\pageref{firstpage}--\pageref{lastpage}}

\maketitle
\label{firstpage}

\begin{abstract}
  
  We have investigated past measurements of the local supermassive black hole
  mass density, correcting for hitherto unknown dependencies on the Hubble
  constant, which, in some cases, had led to an underestimation of the mass
  density by factors of $\sim$2.  Correcting for this, we note that the
  majority of (but not all) past studies yield a local supermassive black hole
  mass density that is consistent with the range 4.4--5.9 $\times10^5 f(H_0)
  M_{\odot}$ Mpc$^{-3}$ (when using $H_0 = 70$ km s$^{-1}$ Mpc$^{-1}$).
  In addition, we address a number of ways in which these past estimates can
  be further developed.  In particular, we tabulate realistic bulge-to-total
  flux ratios which can be used to estimate the luminosity of bulges and 
  subsequently their central black hole masses. 

\end{abstract}


\section{Introduction}

Assuming that the dark mass concentrations at the centres of both elliptical
galaxies and the bulges of disc galaxies are the sleeping engines that powered
past quasar activity (e.g.\ Miller 2006; Brand et al.\ 2005, and references
therein), then the local mass density of such quiescent supermassive black
holes (SMBHs), $\rho_{\rm bh,0}$, can be used to constrain models of quasar
formation and growth (e.g.\ Haehnelt \& Kauffmann 2001; Bromley, Sommerville
\& Fabian 2004; Yu \& Lu 2004; Hopkins, Richards \& Hernquist 2006).
After factoring in potential SMBH mass-energy losses due to gravitational
radiation (Ciotti \& van Albada 2001; Yu \& Tremaine 2002; Menou \& Haiman
2004) and the possibility of lost mass from either ``three-body'' 
SMBH slingshot ejection (e.g.\ Volonteri,
Haardt \& Madau 2003; Hoffman \& Loeb 2007) or explusion 
via gravitational radiation recoil 
(e.g.\ Merritt et al.\ 2004; Libeskind et al.\ 2006) 
$\rho_{\rm bh,0}$ helps constrain the amount of material to 
explain past quasar flux.

Together with the quasar luminosity function (e.g.\ Hopkins et al.\ 2005, and
references therein) integrated over time, $\rho_{\rm bh,0}$ can also
constrain the average efficiency at which matter is converted to radiation 
as it falls onto a SMBH (e.g.\ Ciotti, Haiman \& Ostriker 2001; Elvis et al.\ 
2002; Ferrarese 2002; Yu \& Tremaine 2002; Fabian 2003; Marconi et al.\ 2004;
Merloni 2004; Shankar et al.\ 2004; Yu \& Lu 2004).
This can in turn tell us about the rotation of SMBHs.  For example, a
non-rotating Schwarzschild black hole is expected to have an efficiency of 5.4
per cent while a maximally rotating Kerr black hole may have an efficiency as
great as 37 per cent (Thorne 1974; Hasinger 2005).  Radiative efficiencies are
typically reported to range around 10$-$15 per cent
but values as high as $\sim$ 30$-$37 per cent are also sometimes reported
(e.g.\ Gallo et al.\ 2004; Crummy et al.\ 2006; Wang et al.\ 2006).
%
%

For the above reasons it is of interest to accurately determine $\rho_{\rm bh,0}$. 
In an attempt to help explain some of the differences between previously
reported values (Table~\ref{Tab_Comp}), we will discuss a
number of corrections and adjustments that could be made to past estimates. 
We focus on how estimates of $\rho_{\rm bh,0}$ depend on the Hubble constant,
and how past measurements which have not fully taken this into account are
affected --- sometimes changing by factors of 2 or more.
We shall refer to these revised estimates of $\rho_{\rm bh,0}$ (Section~\ref{SecHub})
as our ``$h$-corrected values''.   In addition we raise a number of other 
points pertaining to the accurate estimation of the local SMBH mass 
density, mostly addressing the issue of recovering the host bulge 
luminosity before converting this into a SMBH mass. 
Section~\ref{SecSum} provides a summary.

\begin{table*}
 \centering
 \begin{minipage}{125mm}
\caption{Local SMBH mass density estimates. 
The factor $h^3_{70} = [H_0/(70$ km s$^{-1}$ Mpc$^{-1})]^3$ 
is appropriate for the Graham et al.\ (2007) study 
because the $M_{\rm bh}$--$n$ relation they used is independent of the Hubble
constant.  The majority of the densities from other papers have been
transformed to $H_0=70$ km s$^{-1}$ Mpc$^{-1}$ using $h^2$ rather than
$h^3$, as indicated in each paper.  However, as 
shown in Table~\ref{Tab_best}, this is not always appropriate. 
\label{Tab_Comp}
}
\begin{tabular}{@{}lccc@{}}
\hline
Study &  $\rho_{\rm bh,0}$ (E/S0)             &  $\rho_{\rm bh,0}$ (Sp)            & $\rho_{\rm bh,0}$ (total) \\
      &  $h^2_{70}10^5 M_{\odot}$ Mpc$^{-3}$   & $h^2_{70}10^5 M_{\odot}$ Mpc$^{-3}$ & $h^2_{70}10^5 M_{\odot}$ Mpc$^{-3}$ \\
\hline
Graham et al.\ (2007)                          & $(3.46\pm 1.16) h_{70}$  & $(0.95\pm 0.49) h_{70}$ & $(4.41\pm 1.67) h_{70}$ \\
Wyithe (2006)                                 &       ...            &       ...           & $2.28\pm0.44$       \\ 
Fukugita \& Peebles (2004)\footnote{See their equation~75.}  &  $(3.4^{+3.4}_{-1.7})h^{-1}_{70}$  &  $(1.7^{+1.7}_{-0.8})h^{-1}_{70}$  &  $(5.1^{+3.8}_{-1.9})h^{-1}_{70}$  \\ 
Marconi et al.\ (2004)                        &       3.3            &      1.3            & $4.6^{+1.9}_{-1.4}$ \\
Shankar et al.\ (2004)\footnote{Based on their Section~3.2.}         &  $3.1^{+0.9}_{-0.8}$  & $1.1^{+0.5}_{-0.5}$ & $4.2^{+1.1}_{-1.1}$ \\
Shankar et al.\ (2004)\footnote{Based on their Section~3.4.}         &  $3.0^{+1.0}_{-0.6}$  & $1.2^{+0.4}_{-0.2}$ & $4.2^{+1.1}_{-0.6}$ \\
McLure \& Dunlop (2004)                       & $2.8\pm0.4$          &      ...            &    ...              \\
Wyithe \& Loeb (2003)                         &       ...            &      ...            & $2.2^{+3.9}_{-1.4}$ \\
Aller \& Richstone (2002)\footnote{Taken from their Table~2.}               & $1.8\pm0.6$          & $0.6\pm0.5$         & $2.4\pm0.8$         \\
Yu \& Tremaine (2002)\footnote{Based on their ($M_{\rm bh}$--$\sigma$)-derived mass function.}                         & $2.0\pm0.2$        & $0.9\pm0.2$       & $2.9\pm0.4$ \\
Merritt \& Ferrarese (2001)\footnote{See also Ferrarese (2002).}                   &        ...           &      ...            &    $4.6h^{-1}_{70}$   \\
Salucci et al.\ (1999)                        &      6.2             &      2.0            &    8.2              \\
\hline
\end{tabular}
\end{minipage}
\end{table*}

\section{Illustrative studies} \label{SecHub}

The nature of the hidden or over-looked dependencies on the Hubble constant
are endemic to most past estimates of $\rho_{\rm bh,0}$.  It is therefore
necessary to only look at a couple of representative case studies in detail,
and provide the revised estimates from other studies in tabular form
(Table~\ref{Tab_best}). 
In what follows, we have chosen two interesting and well written studies.

\subsection{Case study 1}

Our first example is the analysis by Aller \& Richstone (2002, hereafter
AR02), who used the $L$--$\sigma$ relation (e.g.\ Faber \& Jackson 1976) to
convert luminosities into velocity dispersions and then applied the $M_{\rm
  bh}$--$\sigma$ relation (Ferrarese \& Merritt 2000; Gebhardt et al.\ 2000)
to obtain a histogram of SMBH masses. 
We have identified two areas for
improvement pertaining to the treatment of the Hubble constant.

The first is in regard to their adjustment of the Tremaine et al.\
(2002) $M_{\rm bh}$--$\sigma$ relation 
for what they referred to as a 
correction from $h=0.8$ to $h=1$, where $h=H_0/(100 $ km s$^{-1}$ Mpc$^{-1})$.  
Had all, or at least the majority 
of, the SMBH masses used to construct the Tremaine et al.\ 
relation been obtained with distances that
depended on an assumed Hubble constant of 80 km s$^{-1}$ Mpc$^{-1}$,
then it would be appropriate to multiply (decrease) the SMBH masses by
a factor of 0.8 to make the $M_{\rm bh}$--$\sigma$ 
relation consistent with a Hubble constant 
of 100 km s$^{-1}$ Mpc$^{-1}$ (AR02, their equation~23).
However, only five of the 31 galaxies used to construct the Tremaine et
al.\ relation had distances, and thus SMBH masses, derived using a
Hubble constant of 80 km s$^{-1}$ Mpc$^{-1}$; most galaxies had their
distances obtained using surface brightness fluctuations (Tonry et al.\ 2001).  
Removal of the five galaxies with distances obtained using $H_0=80$ km s$^{-1}$
Mpc$^{-1}$ changes neither the slope nor intercept of the Tremaine et al.\
$M_{\rm bh}$--$\sigma$ relation by more than 0.01. 
This relation is therefore effectively 
independent of the Hubble constant and need not be adjusted.
Consequently, the total SMBH mass density in AR02 should be
25 per cent higher and scale with $h^3$ rather than $h^2$.
%
Similarly, the $h$-correction in Yu \& Tremaine (2002, after their equation~6)
which was applied to Tremaine et al.'s (2002) $M_{\rm bh}$--$\sigma$ relation
should also not have been applied.

The SMBH masses that were computed by AR02 are 
dependent on their adopted Hubble constant for a second reason: their
SMBH mass estimates were derived from absolute magnitudes which depend on $h$.  
This SMBH dependence 
on $h$ can be seen in the $\chi$ term which appears in their equation~25, 
and which is defined in their equation~22.  
Removing the aforementioned factor $0.8/h$ from their equation~25, 
one has $M_{\rm bh} \propto h^{-4.02\times 5/7.7} \propto h^{-2.61}$. 
That is, their $M_*$ term varies with $h^{-2.61}$.  
This has apparently gone over-looked in the literature to date. 

Now, AR02's equation~24 for the SMBH mass function, which is 
in units of $M_{\rm bh}$ (rather than $\log M_{\rm bh}$) 
and which depends on $M_*$, can be written as 
\begin{equation}
\frac{{\rm d}N}{{\rm d}M_{\rm bh}} = \frac{h^3 \phi_*}{M_* h^{-2.61}} 
 \left( \frac{M_{\rm bh}}{M_*h^{-2.61}} \right)^{\alpha}
{\rm exp}\left[ - \left( \frac{M_{\rm bh}}{M_*h^{-2.61}} \right)^{\beta} \right]. 
\end{equation}
The expression for the SMBH mass density is thus
\begin{eqnarray}
\rho_{\rm bh}  &=&  \int_{M_{\rm min}}^{M_{\rm max}} M_{\rm bh}
                    \frac{{\rm d}N}{{\rm d}M_{\rm bh}} \hskip3pt {\rm d}M_{\rm bh} \nonumber \\
               &=&  h^3 \phi_* \frac{M_*h^{-2.61}}{\beta}  \{ 
   \gamma \left[ \frac{\alpha+2}{\beta}, \left( \frac{M_{\rm max}}{M_*h^{-2.61}}\right)^{\beta} \right] \nonumber \\
 && - \gamma \left[ \frac{\alpha+2}{\beta}, \left( \frac{M_{\rm
                    min}}{M_*h^{-2.61}}\right)^{\beta} \right] \}, 
\label{FofH}
\end{eqnarray}
where $\gamma (a,x)$ is the incomplete gamma function
(e.g.\ Press et al.\ 1992) defined by
\begin{equation}
\gamma (a,x)=\int ^{x}_{0} {\rm e}^{-t}t^{a-1}{\rm d}t.
\end{equation}
The SMBH mass density in AR02 therefore actually varies with
$h^{3-2.61} = h^{0.39}$ (not $h^2$ as given in AR02) {\it and it also varies} with an additional complicated
dependence on $h$ which is tied up in the gamma functions above.  Correcting
AR02's SMBH mass density to $H_0 = 70$ km s$^{-1}$ Mpc$^{-1}$, and integrating
down to a minimum mass of $10^6 M_{\odot}$, as they did, 
one obtains a value of $\rho_{\rm bh} = 5.9\times10^5 M_{\odot}$ Mpc$^{-3}$.
This value is $\sim$2.5 times larger than what AR02 find 
when $h=0.7$ (see their equation~28).

This type of correction is again not unique to the analysis in AR02, for
example, McLure \& Dunlop's (2004) estimate of $\rho_{\rm bh}$ for E/S0s,
increases from 2.8 $\times10^5 M_{\odot}$ Mpc$^{-3}$ to 4.8$\times10^5
M_{\odot}$ Mpc$^{-3}$ (for $h=0.7$, see Graham et al.\ 2007, their Section~4).
In general, all SMBH mass functions which have been derived from $h$-dependent
galaxy luminosities will depend on $h$ in a similar fashion, although the
above factor of 2.61 may vary from paper to paper (see Table~\ref{Tab_best}).

\subsubsection{Related issues} \label{SecLum}


Ignoring the above mentioned dependencies on the Hubble constant for 
the moment, it is expected that the SMBH masses 
in AR02 are too high at the low-mass end because of a)
the way they converted disc galaxy magnitudes into bulge magnitudes,
and b) the way they assigned a velocity dispersion to these
magnitudes.
The average bulge-to-disc ($B/D$) luminosity ratios which AR02 assigned to
their early- and late-type spiral galaxies, and also lenticular galaxies, came
from the $R^{1/4}$-bulge plus exponential-disc decompositions in Simien \& de
Vaucouleurs (1986).  Due to Simien \& de Vaucouleurs use of the $R^{1/4}$
model to describe bulges which are better matched with an $R^{1/n}$ profile
having $n < \sim3$, and often around 1 (e.g.\ Andredakis \& Saunders 1994; de
Jong 1996; Balcells et al.\ 2003), too much flux has been assigned to the
bulges of their disc galaxies.
We have derived the mean bulge-to-total ratios from various studies and show
the results in Table~\ref{Tab_BD}.  On average, the bulge luminosities used by
AR02 will be $\sim$2 times too bright 
%
%
and thus their estimate of the SMBH mass in the bulges of disc galaxies will
be high by a factor of $\sim$2.  Correcting for this would result in a 12.5
per cent reduction to their {\it total} SMBH mass density, giving a value of
5.2 $\times10^5 M_{\odot}$ Mpc$^{-3} (H_0 = 70$ km s$^{-1}$ Mpc$^{-1}$).
Using Allen et al.'s (2006) S\'ersic-bulge $+$ exponential disc decompositions of 10,095
galaxies, we plan to apply the $M{\rm bh}$-$L$ relation from Graham (2007) to 
obtain a new measurement of $\rho_{\rm bh,0}$. 

\begin{table}
\begin{minipage}{85mm}
\caption{Mean $\pm$ standard deviation of the (bulge minus
galaxy) magnitude and bulge-to-galaxy flux ratios, $B/T$, 
derived from literature data. 
(A more complete summary, using other literature data, and a variety
of optical and near-infrared passbands will be presented in 
Graham \& Worley 2007, in prep.) 
\label{Tab_BD}
}
\begin{tabular}{@{}lccc@{}}
\hline
\hline
 & S0/S0a & Sa,Sab,Sb      & $\geq$Sc \\
\hline
\multicolumn{4}{c}{$R^{1/4}$-bulge plus Exponential-disc} \\
($m_{\rm bulge} - m_{\rm tot}$)\footnote{$B$-band data from Simien \& de Vaucouleurs 1986.}  &  $0.61\pm0.32$  &  $1.37\pm0.68$  &  $3.22\pm0.99$  \\
($B/T$)$^a$                                                               &  $0.59\pm0.16$  & $0.34\pm0.20$ & $0.07\pm0.06$ \\
[2pt]
($m_{\rm bulge} - m_{\rm tot}$)\footnote{$B$-band data from de Jong 1996.}                    &   ...           &  $1.67\pm1.06$  &  $3.18\pm1.41$  \\
($B/T$)$^b$                                                               &   ...           & $0.32\pm0.26$ & $0.11\pm0.11$ \\
[5pt]
\multicolumn{4}{c}{Exponential-bulge plus Exponential-disc} \\
($m_{\rm bulge} - m_{\rm tot})^b$                                         &   ...           &  $2.70\pm1.18$  &  $4.08\pm1.03$  \\
($B/T$)$^b$                                                               &   ...           & $0.12\pm0.10$ & $0.04\pm0.03$ \\
[5pt]
\multicolumn{4}{c}{$R^{1/n}$-bulge plus Exponential-disc}   \\
($m_{\rm bulge} - m_{\rm tot}$)\footnote{$B$-band data from Graham 2003.}                    &   ...           &  $2.36\pm1.06$  &  $4.21\pm1.06$  \\
($B/T$)$^c$                                                               &   ...           & $0.17\pm0.09$ & $0.03\pm0.03$ \\
[2pt]
($B/T$)\footnote{$K$-band data from Balcells, Graham \& Peletier 2007.}    & $0.25\pm0.09$   &     ...         &      ...        \\
($B/T$)\footnote{$K$-band data from Laurikainen, Salo \& Buta 2005.}      & $0.24\pm0.11$ &     ...         &      ...        \\
\hline
\end{tabular}
\end{minipage}
\end{table}

Regarding the conversion of these overly-bright bulge magnitudes to velocity
dispersions, the logarithmic slope of the $L$--$\sigma$ relation is known to
be shallower at fainter luminosities (e.g.\ Tonry 1981, Held et al.\ 1992),
and for magnitudes below $M_B \sim -19.5\pm1$ mag the slope is approximately
2 (de Rijcke et al.\ 2005; Matkovi\'c \& Guzm\'an 2005), compared to a value
of four for the more luminous spheroids (Faber \& Jackson 1976; their
figure~16).  The use of a constant slope of 3 by AR02 would have therefore
systematically under-estimated the velocity dispersion as one progresses to
fainter magnitudes, and over-estimated the velocity dispersion in the larger
spheroids.
%
%
Without performing a full re-analysis of their data, the overall corrective
term is unknown. 
In passing we note that the non-linear nature of the $L$--$\sigma$ relation
complicates AR02's prediction of the parameter $\beta$ shown in
their equation~10.

\subsection{Case study 2}

Our second example is the analysis in Shankar et al.\ 
(2004, their section 3.1 \& 3.2), who used the $M_{\rm bh}$--$L$ relation,
in addition to the $M_{\rm bh}$--$\sigma$ relation, to
estimate $\rho_{\rm bh}$ from various luminosity functions.  
Their equation~1 for predicting SMBH masses from luminosities was also
obtained under the (false) assumption that the SMBH masses which define this
relation are dependent on the Hubble constant.  They modified the 
$M_{\rm bh}$--$L$ relation from McLure \& Dunlop (2002; their equation~6) 
which had originally been (correctly) constructed with no $H_0$-adjustment 
to the black hole masses. 
The equation in Shankar et al.\ therefore requires that $\log (70/50)$ be subtracted 
from the left hand side\footnote{Equation~1 from McLure \& Dunlop (2004) 
should read $\log(M_{\rm bh}/M_{\sun}) = 1.25\log(L_K/L_{\sun})-5.53$, when 
using $M_{K,\sun}=3.28$ mag and $H_0 = 70$ km s$^{-1}$ Mpc$^{-1}$.
While relevant to the $\rho_{\rm bh}$ adjustments made in Table~\ref{Tab_best}, 
this is perhaps a moot point given that the galaxy distances and thus absolute 
magnitudes used to construct that equation are known independently of the
Hubble constant, just as the SMBH masses are, see section 3.1 of Graham 2007.}.
Correcting this 
results in a 40 per cent increase to their ($M_{\rm bh}$--$L$)-estimated SMBH
masses and thus a 40 per cent increase in their value of $\rho_{\rm bh}$.
Their $h$-corrected value for $\rho_{\rm bh,total}$ is $(5.9 \pm 1.5)
\times10^5 M_{\odot}$ Mpc$^{-3} (h=0.7)$, in perfect agreement with AR02's
$h$-corrected value.
Similarly, their equation~3 should not contain the factor 80/$H_0$, and so
their ($M_{\rm bh}$--$\sigma$)-estimated SMBH masses, and thus their ($M_{\rm
  bh}$--$\sigma$)-derived $\rho_{\rm bh,0}$, needs to be increased by 14 per
cent when using their adopted value of $h=0.7$.

Shankar et al.'s equation~4 for the SMBH mass function, which has $\phi_*$ 
in units of $\log M_{\rm bh}$ (rather than $M_{\rm bh}$) can be written as 
\begin{equation}
\frac{{\rm d}N}{{\rm d}\log M_{\rm bh}} = h^3 \phi_* 
 \left( \frac{M_{\rm bh}}{M_*h^{-2.5}} \right)^{\alpha + 1}
{\rm exp}\left[ - \left( \frac{M_{\rm bh}}{M_*h^{-2.5}} \right)^{\beta}
 \right], 
\end{equation}
where the exponent $-2.5$ comes from their equation~1 which was used to transform
magnitudes $M_R$ into SMBH masses using $\log M_{\rm bh} \propto -0.5(M_R + 5\log
h)$. The expression for the SMBH mass density is thus
\begin{eqnarray}
\rho_{\rm bh}  &=&  \int_{M_{\rm min}}^{M_{\rm max}} M_{\rm bh}
                    \frac{{\rm d}N}{{\rm d}\log M_{\rm bh}} \hskip3pt {\rm
                    d}\log M_{\rm bh} \nonumber \\
               &=&  h^3 \phi_* \frac{M_*h^{-2.5}}{\beta (\ln 10)}  \{ 
   \gamma \left[ \frac{\alpha+2}{\beta}, \left( \frac{M_{\rm max}}{M_*h^{-2.5}}\right)^{\beta} \right] \nonumber \\
 && - \gamma \left[ \frac{\alpha+2}{\beta}, \left( \frac{M_{\rm
                    min}}{M_*h^{-2.5}}\right)^{\beta} \right] \}. 
\end{eqnarray}
A similar parameterisation of the SMBH mass function could be made for the
data in 
McLure \& Dunlop (2004) and Marconi et al.\ (2004), except for the latter
study the exponent would be $-2.26$ rather than $-2.5$ (see their equation~10).
This full dependency on the Hubble constant was not included in Tundo et al.'s
(2007) reanalysis of these works.  Their $M_{\rm bh} \propto L^{1.30}$
relation is also considerably steeper than the new expression $M_{\rm bh}
\propto L^{0.93}$ reported by Graham (2007) and it predicts notably
larger SMBH masses for galaxies more luminous than $M_R \sim -21$ mag.

In Table~\ref{Tab_best} we provide updated values of $\rho_{\rm bh}$ and, importantly,
show their dependence on the Hubble constant.  While some estimates of 
$\rho_{\rm bh,0}$ appear not to have changed from Table~\ref{Tab_Comp}, 
one should note that the quoted dependence on $h$ may have changed, which is 
of course of importance if $H_0 \neq 70$ km s$^{-1}$ Mpc$^{-1}$.

\begin{table*}
 \centering
 \begin{minipage}{150mm}
\caption{Modification of Table~\ref{Tab_Comp}.  
Here, the local SMBH mass density estimates have been fully corrected
for their dependence on $h$, and transformed to 
$H_0=70$ km s$^{-1}$ Mpc$^{-1}$.  
While some estimates of
$\rho_{\rm bh,0}$ appear not to have changed from Table~\ref{Tab_Comp},
one should note that the quoted dependence on $h$ may have changed. 
The term $f(h)$ is used to 
denote that a more complicated dependence on $h$ exists 
and needs to be taken into account if one is to transform these values to a 
different Hubble constant (see, e.g.\ equation~\ref{FofH}). 
\label{Tab_best}
}
\begin{tabular}{@{}lcccc@{}}
\hline
Study &  Method   &   $\rho_{\rm bh,0}$ (E/S0)      &  $\rho_{\rm bh,0}$ (Sp)       &  $\rho_{\rm bh,0}$ (total) \\
      &           &   $10^5 M_{\odot}$ Mpc$^{-3}$   &  $10^5 M_{\odot}$ Mpc$^{-3}$  &  $10^5 M_{\odot}$ Mpc$^{-3}$ \\
\hline
Graham et al.\ (2007)  &  $M_{\rm bh}$--$n$         &  $(3.46\pm 1.16) h^3_{70}$      &  $(0.95\pm 0.49) h^3_{70}$  &  $(4.41\pm 1.67) h^3_{70}$ \\
Wyithe (2006)         &  $M_{\rm bh}$--$\sigma$    &       ...                     &       ...                 &  $(1.98\pm0.38) h^3_{70}$       \\ 
Fukugita \& Peebles (2004)  &  $\rho_{\rm spheroid}$  &  $(3.4^{+3.4}_{-1.7})h_{70}$  &  $(1.7^{+1.7}_{-0.8})h_{70}$  &  $(5.1^{+3.8}_{-1.9})h_{70}$ \\ 
Marconi et al.\ (2004) &  $M_{\rm bh}$--($L,\sigma)$ &  $3.3h^{0.74}_{70}f(h)$      &  $1.3h^{0.74}_{70}f(h)$   &  $(4.6^{+1.9}_{-1.4})h^{0.74}_{70}f(h)$  \\
Shankar et al.\ (2004)  &  $M_{\rm bh}$--$L$   & $(4.3^{+1.3}_{-1.1})h^{0.5}_{70}f(h)$  &  $(1.5^{+0.7}_{-0.7})h^{0.5}_{70}f(h)$ &  $(5.9^{+1.5}_{-1.5})h^{0.5}_{70}f(h)$ \\
Shankar et al.\ (2004)  &  $M_{\rm bh}$--$\sigma$  &  $(3.4^{+1.1}_{-0.7})h^3_{70}$  &  $(1.4^{+0.5}_{-0.3})h^3_{70}$ &  $(4.8^{+1.2}_{-0.8})h^3_{70}$ \\
McLure \& Dunlop (2004) &  $M_{\rm bh}$--$L$        & $(4.8\pm0.7)h^{0.5}_{70}f(h)$ &      ...                &    ...              \\
Wyithe \& Loeb (2003)   &  $M_{\rm bh}$--$\sigma$   &         ...                   &      ...             &  $(2.1^{+3.4}_{-1.3})h^3_{70}$ \\
Aller \& Richstone (2002)  &  $M_{\rm bh}$--$\sigma$  &  $(4.5\pm1.5) h^{0.39}_{70}f(h)$  &  $(1.4\pm1.3)h^{0.39}_{70}f(h)$  &  $(5.9\pm2.0)h^{0.39}_{70}f(h)$ \\
Yu \& Tremaine (2002)   &  $M_{\rm bh}$--$\sigma$   &  $(2.0\pm0.2)h^3_{70}$        &  $(0.9\pm0.2)h^3_{70}$    &  $(2.9\pm0.4)h^3_{70}$    \\
Merritt \& Ferrarese (2001) &  $\rho_{\rm spheroid}$  &        ...                 &      ...                  &   $4.6h_{70}$       \\
Salucci et al.\ (1999)      &  $\rho_{\rm spheroid}$  &   $6.2h^{2}_{70}$        &  $2.0h^{2}_{70}$        &  $8.2h^{2}_{70}$  \\
\hline
\end{tabular}
\end{minipage}
\end{table*}


\subsubsection{Related issues} \label{subsec}

Many studies have assumed the universal existence 
of $R^{1/4}$ light-profiles when obtaining their total galaxy
magnitudes, and have thus introduced a systematic bias into their
luminosity-derived SMBH mass function.  For instance, a light-profile shape
dependent --- and therefore luminosity dependent (e.g.\ Graham \& Guzm\'an
2003, their figure~10 and references therein) --- magnitude correction (Graham
et al.\ 2005) is applicable to the SDSS Petrosian magnitudes which Shankar et
al.\ used.  Adding $-0.2$ mag to the Petrosian magnitudes, in an effort to
recover the total galaxy magnitude,
is only applicable if every galaxy has an $R^{1/4}$ light-profile.  However, a
range of profile shapes has long been known to exist (e.g.\ Davies et al.\ 
1988; Caon et al.\ 1993) and is such that a smaller/greater correction for
missed flux needs to be applied to the Petrosian magnitudes of galaxies
less/more luminous than $M_B \sim -21$ mag (Kormendy \& Djorgovski 1989).
 Similarly, a light-profile shape (and outermost sampled radius) dependent
 magnitude correction (Graham \& Driver 2005, their figure~10) is required for
 recovering total magnitudes from Kron magnitudes.  Indeed, 
 half a galaxy's flux may be missed using Kron magnitudes 
 (Andreon 2002; Bernstein, Freedman, \& Madore 2002; Benitez et al.\ 2004). 


\section{Summary} \label{SecSum}

Table~\ref{Tab_best} shows our ``$h$-corrected'' $\rho_{\rm bh,0}$ values.  It
should be noted that the $h$-dependent corrections we have detailed effect not
only the value of $\rho_{\rm bh,0}$ but also the SMBH mass functions from
which these values are typically derived.  The related issues we have raised
in sections~\ref{SecLum} and \ref{subsec} that pertain to the luminosity of
the host spheroid have not been included in Table~\ref{Tab_best}.
%
%
The use of $h$-independent S\'ersic indices and velocity dispersions for
constructing the SMBH mass function and mass density results in a purely $h^3$
dependence for $\rho_{\rm bh}$.  This is because the SMBH masses that are
involved are derived from relations which themselves do not depend on any
assumed Hubble constant.

We (tentatively) identify previously missed agreements on the value of 
$\rho_{\rm bh,0}$.  For example,
%
%
AR02's corrected value\footnote{Computing AR02's disc galaxy's bulge
  luminosities with more realistic $B/T$ flux ratios leads to a 12.5 per cent
  reduction in $\rho_{\rm bh,0}$, giving a value of $(5.2 \pm1.8) \times10^5
  M_{\odot}$ Mpc$^{-3}$ (for $h=0.7$).}  
of $(5.9 \pm2.0) \times10^5 M_{\odot}$ Mpc$^{-3}$ (for $h=0.7$) is now 
in good agreement with Merritt \& Ferrarese's (2001) ($M_{\rm 
  bh}$--$\sigma$)-derived measurement of $4.6 \times10^5 h^3_{70} M_{\odot}$
Mpc$^{-3}$.  
In fact, 
a (near) consensus on the local SMBH mass density now exists.
%
%
%
The $M_{\rm bh}$--$L$ based studies are seen to agree with each other and with
recent studies which have used a mean SMBH-to-spheroid mass ratio convolved with the
local spheroid mass density.  The $M_{\rm bh}$--$n$ based study (Graham et
al.\ 2007) is also seen to agree with both of 
these types of analysis, with the optimal (total) SMBH mass densities ranging
from 4.6--5.9 $\times10^5 M_{\odot}$ Mpc$^{-3} (h=0.7)$ for all three types
of analysis.  Furthermore, some of the $h$-corrected $M_{\rm bh}$--$\sigma$ based studies
(Aller \& Richstone 2002; Marconi et al.\ 2004; Shankar et al.\ 2004) also
provide consistent results with this range.  The two exceptions are
the noticeably lower values of $(2.9 \pm 0.4) \times10^5 h^3_{70} M_{\odot}$
Mpc$^{-3}$ (Yu \& Tremaine 2002) and $(2.0 \pm 0.4) \times10^5 h^3_{70}
M_{\odot}$ Mpc$^{-3}$ (Wyithe 2006)\footnote{Correcting Wyithe's (2006)
estimate of $\rho_{\rm bh,0}$ for the intrinsic scatter in the
$M_{\rm bh}$--$\sigma$ relation (see equation~12 in Yu \& Tremaine
2002) would only increase it by a factor of 1.1--1.2.}.  
%

Excluding galaxies without `secure' SMBH mass determinations, Marconi et al.\ 
(2004) derived and used an $M_{\rm bh}$--$\sigma$ relation with a 0.17 dex higher
zero-point (and 0.09 steeper slope) than used by Yu \& Tremaine (2002).  
This accounts for their different ($M_{\rm bh}$--$\sigma$)-derived values of
$\rho_{\rm bh,0}$. 
%
%
It is also worth noting that {\it if} the local sample of $\sim$30 galaxies
with direct SMBH mass measurements have low luminosities with respect to the
greater population at any given velocity dispersion (Yu \& Tremaine 2002;
Bernardi et al.\ 2007; Tundo et al.\ 2007; Lauer et al.\ 2007), then the
$M_{\rm bh}$--$L$ relation will over-predict $\rho_{\rm bh,0}$. However, as
noted by Graham (2007, his Appendix), until accurate bulge/disc decompositions
are available for the greater population, and corrections for dust attenuation
in the bulges of disc galaxies are addressed (Driver et al.\ 2007), this
remains uncertain.

\section{acknowledgments}

We are grateful to Paul Allen and Joe Liske for proof reading 
this work prior to an October 02, 2006 submission as the 
Appendix of a different paper.  We also thank 
A.Marconi, 
D.Merritt, 
F.Shankar 
and 
Q.Yu
for their comments.

\label{lastpage}
\end{document}